\documentclass[prl,showpacs,twocolumn]{revtex4}
\usepackage{amssymb,graphicx,afterpage}
\begin{document}
\draft

\noindent
{\bf Kawasaki, Araki and Tanaka Reply:} In the preceding Comment 
\cite{Tarjus1}
Sausset and Tarjus (ST) proposed an alternative scenario for the slow 
dynamics in our two-dimensional (2D) polydisperse colloidal liquids 
\cite{Kawasaki}, based on the frustration-limited domain 
theory \cite{Tarjus2} which focuses on self-generated frustration 
in the order parameter itself as in \cite{NelsonB}. 
ST claimed that sufficiently polydisperse hexatic order is not 
space-filling, so it is a 2D analog of the icosahedron. 
We agree with the former, but the latter seems to be subtle 
due to the lack of the uniformity of frustration. 
An exact 2D analog may be hexatic 
ordering on a surface of incommensurate constant curvature 
in the sense that in both cases frustration is `uniform' 
\cite{Tarjus2,NelsonB}. 
We regard the same phenomenon as 
random-field effects on (quasi-)long-range 
crystalline ordering \cite{Kawasaki,Tanaka1,Tanaka2}. 
Since we described our thoughts on 
the differences between the two approaches in detail 
in \cite{Tanaka2}, we do not repeat it here. 

First we show the analysis proposed by ST in Fig. 1(a). 
Their function $L^\ast \sim B[(\phi-\phi_I)/\phi_I]^x+C$ fits 
reasonably well to our data. 
Here $\phi$ is the volume fraction of colloids and 
$\phi_I$ is $\phi$ at the hexatic ordering for polydispersity $\Delta=0$ \% 
($\phi_I \sim 0.57$) [Fig. 1(b)]. 
The fitting yields $x \sim 3$, consistent with 
the suggestion of ST \cite{Tarjus1}. 
$x$ is suggested to be 
related to the correlation length exponent of the unfrustrated system 
\cite{Tarjus2}: For the present case (2D hexatic ordering), 
$\xi_6 \sim e^{h[(\phi_I-\phi)/\phi]^{-1/2}}$ 
\cite{NelsonB}. The physical meaning of $x \sim 3$ needs to be clarified 
along this line.  
Since our model predicts the divergence of $\xi$ toward $\phi_0$ 
whereas their model predicts the absence of any such singularity, 
the difference between the two predictions 
should more evidently appear near $\phi_0$. 
So we made simulations at $\phi=0.64$ 
for a system of 16384 particles [see the points 
in the yellow (shaded) circle in Fig. 1(a)]. 
Unfortunately, the difference is too small to draw any conclusions. 

\begin{figure}[b]
\begin{center}
\includegraphics[width=8.0cm]{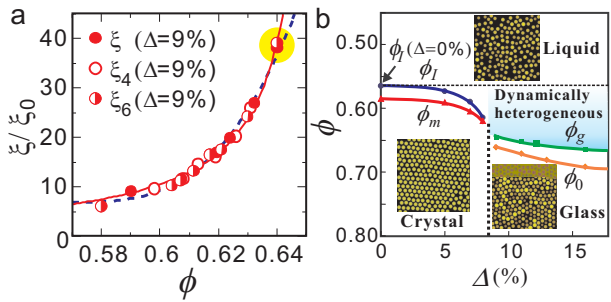}
\end{center}
\caption{(color online). 
(a) $\phi$ dependence of the scaled correlation length $\xi/\xi_0$ 
for $\Delta=9$ \%. 
Solid and dashed curve are the fittings of  
$(\phi_0/\phi-1)^{-1}$ and $(\phi-\phi_I)^{3}$, respectively. 
(b) State diagram of polydisperse colloidal liquids in the 
$\phi$-$\Delta$ plane. 
}
\label{figure1}
\end{figure} 

Here we mention the work of Santen and Krauth \cite{Krauth}, 
which demonstrate that there is no ideal glass transition point $\rho_G$ 
for $\Delta \sim 50$ \% (in our definition of $\Delta$ \cite{Kawasaki}). 
They determined $\rho_G$ by fitting the diffusivity $D$ 
by $(\rho_G-\rho)^\alpha$ as $\rho_G= 0.805$. 
This $\rho_G$ corresponds to $\phi_G = \rho_G/\sqrt[3]{2} =0.64$ 
in our notation. 
This fitting function is not the Vogel-Fulcher type, but that for 
mode-coupling theory. 
Thus, $\phi_G$ is the mode-coupling $\phi_C$ and we expect that 
$\phi_0 > \phi_G=\phi_C$. 
In our opinion, thus, what they demonstrated is 
that there is no thermodynamic phase 
transition at the mode coupling $\phi_C$. 
According to our state diagram [Fig. 1(b)], $\phi_G=0.64$ 
for $\Delta \sim 50$ \% may be located far below $\phi_0$. 
Our experiments on 2D driven granular systems \cite{Watanabe} 
also demonstrated 
that for $\Delta=10.7$ \% $\rho_0=0.838$, 
which is higher than $\rho_G$ for $\Delta \sim 50$ \% \cite{Krauth}.

Next we mention our previous simulation study of a system with competing 
orderings \cite{Shintani}. In this case, the underlying crystalline 
order is anti-ferromagnetic and the crystallization is 
of first order. Nevertheless, we observe behavior very similar 
to the present case. 
The basic features of the phase diagram are also very similar 
between the two [compare Fig. 1(b) 
with Fig. 2 of \cite{Shintani}]. These facts seem to support our scenario. 

Finally, we note that our preliminary study on 3D polydisperse 
colloidal liquids indicate that there exists medium-range crystalline 
ordering (fcc or hcp), which is not icosahedral, 
and $\xi \propto (\phi_0/\phi-1)^{-2/3}$, 
consistent with our prediction \cite{Tanaka1}. 
This also supports our scenario. 
At the same time, however, a recent study by Coslovich and Pastore 
favors the scenario of frustration-limited domain theory \cite{Tarjus2} 
rather than ours. Thus, further careful studies are required to settle 
the issue of the role of frustration in the glass transition. 
Such efforts will ultimately lead to 
a clear physical understanding of the glass transition. 

\vspace{0.4cm}
\noindent
Takeshi Kawasaki, Takeaki Araki, and Hajime Tanaka$^\ast$

\noindent
Institute of Industrial Science, University of Tokyo, 
Meguro-ku, Tokyo 153-8505, Japan.

\noindent
$^\ast$tanaka@iis.u-tokyo.ac.jp


\end{document}